%% file: hyperon_asymm.tex
\documentstyle{elsart}
\input{epsf}

\begin{document}

\def\ss{\scriptscriptstyle}
\def\ysub{_{\scriptscriptstyle Y}}
\def\ybarsub{_{\scriptscriptstyle \overline{Y}}}
\def\ysubb{_{\scriptscriptstyle Y - \scriptscriptstyle \overline{Y}}}
\def\xf{x_{\scriptscriptstyle F}}
\def\pt{p_{\scriptscriptstyle T}^2}
\def\pbar{\overline{p}}
\def\Do{D^{0}}
\def\Dobar{\overline{D}^{0}}
\def\Ds{D^+_{\mbox{\footnotesize{s}}}}
\def\Lm{\Lambda^{\, 0}}
\def\Lbar{\overline{\Lambda}^{\, 0}}
\def\Lc{\Lambda_c^{\, +}}
\def\Lcbar{\Lambda_c^{\, -}}
\def\Xim{\Xi^-}
\def\Xibar{\Xi^+}
\def\Om{\Omega^-}
\def\Ombar{\Omega^+}
\def\PYTHIA{\textsc{Py\-thia}}
\def\JETSET{\textsc{Jet\-set}}

\def\etal{{\em et al.}}
\def\PL#1{Phys. Lett. {#1}}
\def\PR#1{Phys. Rev. {#1}}
\def\PRL#1{Phys. Rev. Lett. {#1}}
\def\NP#1{Nucl. Phys. {#1}}
\def\NIM#1{Nucl. Instr. and Methods {#1}}

\begin{frontmatter}

\title{Asymmetries in the production of $\Lm$, $\Xim$,
and $\Om$ hyperons in 500 GeV/$c$ $\pi^-$ -- Nucleon Interactions}
\collab{Fermilab E791 Collaboration}
\input{author.tex}

\begin{abstract}
Using data from Fermilab fixed-target experiment E791, we have measured
particle-antiparticle production asymmetries for $\Lm$, $\Xim$,
and $\Om$ hyperons in $\pi^-$ -- nucleon interactions at 500 GeV/$c$.
The asymmetries are measured as
functions of Feynman-x ($\xf$) and $\pt$ over the ranges
$-0.12\leq\xf\leq 0.12$ and $0\leq\pt\leq 4 (GeV/c)^2$.  We find
substantial asymmetries, even at $\xf$ = 0. We also observe
leading-particle-type asymmetries which qualitatively agree with
theoretical predictions.
\end{abstract}
\end{frontmatter}

   Strange particle production is an important tool for studying how
non-perturbative QCD affects light quark production and hadronization. One
process that involves both production and hadronization is the leading
particle effect. This effect is manifest as an enhancement in the production
rate of particles which have one or more valence quarks in common with an
initial state hadron compared to that of their antiparticles which have fewer
valence quarks in common. This enhancement increases as the momentum of the
produced particle increases, in the direction of the initial hadron with which
the produced particle shares valence quarks. This process has been extensively
studied in charm production in recent years from both
experimental~\cite{charm-exp,sudeshna} and theoretical~\cite{charm-mod}
points of view. The same type of leading particle effects have been seen in
light hadron production~\cite{strange-mod1} and are
expected~\cite{strange-mod}. Other effects, like the associated
production of a kaon and a hyperon, can also contribute to an asymmetry in
hyperon-antihyperon production~\cite{capella1}.

        As a byproduct of our charm program in Fermilab experiment E791, we
collected a large sample of $\Lm-\Lbar$, $\Xim-\Xibar$, and $\Om-\Ombar$
hyperons which we have used to measure the particle-antiparticle production
asymmetries reported here. For hyperons, the range of scaled
longitudinal momentum $\xf$ ($2p_L/\sqrt{s}$) covered by our experiment was
$-0.12 \leq \xf \leq 0.12$. Given E791's negative pion beam and nucleon target,
we expect recombination effects~\cite{strange-mod} to produce different
asymmetries in the beam and target fragmentation regions as the content of
valence quarks in the beam and target hadrons differ.
A growing asymmetry is expected in $\Lm-\Lbar$\ production as
$\left| \xf \right|$ increases in the negative direction towards the target
fragmentation region. A smaller asymmetry (or none at all) is expected for
$\xf>0$, towards the beam fragmentation region, because $\Lm$ and $\Lbar$
each share one valence quark with the incident pion. For $\Xim-\Xibar$
production, a growing asymmetry with $\left| \xf \right|$ is expected in both
regions since the $\Xim$ shares one valence quark with the $\pi^-$ as well as
with the target particles ($p$ and $n$) whereas the $\Xibar$ shares none. No
leading particle effects are expected for $\Om$ or $\Ombar$ as they have no
valence quarks in common with either the target or the beam.

       Measurements of $\Lm-\Lbar$ \cite{accmor1}, $\Xim-\Xibar$, and
$\Om-\Ombar$ \cite{accmor2} production asymmetries in $\pi^-$Cu interactions
at 230 GeV/$c$ have been reported in the literature. Additional measurements
of the $\Lm-\Lbar$ asymmetry can be found in the experiments listed in
\cite{add-evid}, but in general light hadron production asymmetries in
$\pi^-N$ interactions have not been studied extensively.

    Experiment E791 recorded data from 500 GeV/$c$ $\pi^-$ interactions in
five thin foils (one platinum and four diamond) separated by gaps of 1.38 to
1.39 cm. Each foil had a thickness of approximately 0.4$\%$ of a pion
interaction length (0.5 mm for the upstream platinum target, and 1.6 mm for
each of the
carbon targets). The E791 spectrometer \cite{viteum} in the Fermilab Tagged
Photon Laboratory was a large-acceptance, two-magnet spectrometer augmented by
eight planes of multiwire proportional chambers (MWPC) and six planes of
silicon microstrip detectors (SMD) for beam tracking. The magnets
provided a total transverse momentum kick of 512 MeV/$c$. Downstream of the
target there were 17 planes of SMD's for track and vertex reconstruction, 35
drift chamber planes, two MWPC's, two multicell threshold \v{C}erenkov
counters, electromagnetic and hadronic calorimeters (with apertures about 70
by 140 mr), and a muon detector. An important element of the experiment was
its extremely fast data acquisition system~\cite{amato} which was combined
with a very open transverse-energy trigger to record a data sample of
$2 \times 10^{10}$ events. The trigger required that the total
``transverse energy'' (i.e., sum of the products of energy
observed times the tangent of the angle from the target to each calorimeter
segment)
be at least 3 GeV.


    For this analysis we use only interactions in the isoscalar carbon targets
so that our results truly represent a ``nucleon'', that is, the average of
neutrons and protons.
Most $\Lm$'s decay before entering the drift chamber
region (150 cm downstream of the targets) but downstream of the end of the
silicon vertex detectors (50 cm from the targets), while some $\Xim$'s
and $\Om$'s decay in the silicon region.

\begin{figure}[b]
\centering
  \mbox{\epsfxsize=7.0in \epsffile{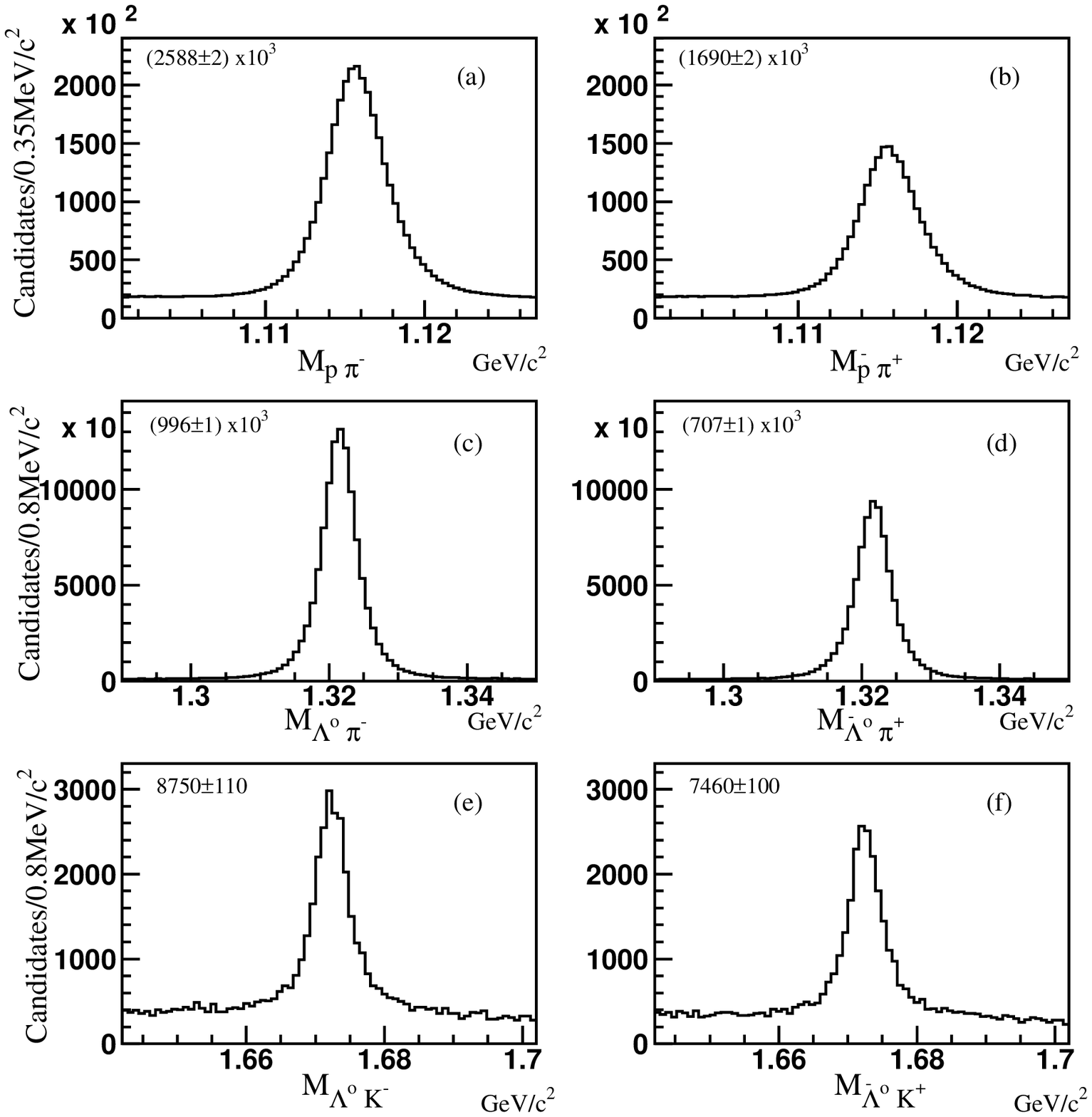}}
  \caption{Effective mass distributions for decay products of hyperons
   in events selected for this study and the corresponding signals with
background subtracted.  Plots correspond to
   (a) $\Lm\rightarrow p\pi^-$;
   (b) $\Lbar\rightarrow \pbar\pi^+$;
   (c) $\Xim\rightarrow\Lm\pi^-$;
   (d) $\Xibar\rightarrow\Lbar\pi^+$;
   (e) $\Om\rightarrow\Lm K^-$
   (f) $\Ombar\rightarrow\Lbar K^+$.
The numbers and errors in the top corner of each plot come from the sum of the
number of background subtracted events in the individual ($\xf,\pt$) bin
fits.}
  \label{fig-masses}
\end{figure}

  Throughout this paper, references to a particle should be taken to include
its antiparticle unless explicitly stated otherwise.
  $\Lm$'s were reconstructed using the $p\pi^-$ decay mode. Proton and $\pi^-$
tracks were required to have a distance of closest approach less than 0.7 cm
at the decay vertex and to have an invariant mass between 1.101 and 1.127
GeV/$c^2$.  In addition, the ratio of the momentum of the proton to that of
the pion was required to be larger than 2.5. The reconstructed $\Lm$ decay
vertex formed by the two tracks was required to be downstream of the last
target but upstream of the first magnet. For the $\Lm$ production study, we
removed $\Lm$'s coming from $\Xim$
decay by requiring that the $\Lm$ candidates have an impact parameter with
respect to the primary vertex of less than 0.3 cm if decaying within the
first 20 cm downstream of the target, and less than 0.4 cm otherwise.
After these cuts, the remaining $\Xim$ contamination was $\approx 1.5\%$,
having
negligible effect on the $\Lm-\Lbar$ asymmetry. The reconstructed mass
distributions for this sample, for each interval of $\xf$ and $\pt$, were
fit using a binned maximum likelihood method with a Gaussian signal plus a
linear background. All bin widths were much larger than the experimental
resolution of the variable binned.
In the fit, the central reconstructed mass values and mass resolutions were
fixed to values obtained from Monte Carlo simulation.
 Contamination from misidentified
$K^0_s\rightarrow\pi^+\pi^-$ decays provided a flat background in these
distributions in the range shown, and produced a negligible effect on the fit
numbers of $\Lm$'s.
The reconstructed mass distributions for the entire sample are shown in
Figs.~\ref{fig-masses}(a) and (b). For the $\Lm-\Lbar$ analysis we use data from
approximately 7$\%$ of the overall sample recorded for the experiment.
The total signal, taken as the sum of
background subtracted signal in each bin,
was 2\,587\,870 $\pm$ 1\,780 $~\Lm$'s and 1\,690\,030 $\pm$ 1\,500 $~\Lbar$'s.

     $\Xim$ were selected via the decay mode $\Xim\rightarrow\Lm\pi^-$ and
$\Om$ via the decay mode $\Om\rightarrow\Lm K^-$.
Beginning with a $\Lm$ candidate, we added a third, distinct track as a possible
pion or kaon daughter.  All three tracks were required to have hits
in the drift chamber region only. For these samples, we removed the requirement
on the $\Lm$ impact parameter.
The invariant mass of the candidate hyperon
(calculated from the known mass and measured momentum of the $\Lm$, as
determined by its two decay tracks, together with the third track) was
required to be between  1.290 and 1.350 GeV/$c^2$ for the $\Xim$ and
between 1.642 and 1.702 GeV/$c^2$ for the $\Om$. The charged hyperon track,
entirely in the SMD region, was required to be reconstructed. Its direction had to
match that determined by its daughter tracks in the drift chamber region to
within 1 mrad and it was required to have a distance of closest approach to
the primary vertex of less than 100 $\mu$m. The $\Xim$ or $\Om$ vertex
(reconstructed from the $\Lm$ and daughter $\pi^-$ or $K^-$) was
required to be downstream of the last SMD plane and upstream of the
$\Lm$ vertex. The former requirement allows us to track the hyperon
in the silicon.
For the $\Om$ sample, the third track was required to have a kaon
signature in the \v{C}erenkov counters and momentum in the range 6-40
GeV/$c$. For this momentum range, the \v{C}erenkov  kaon identification
efficiency was about 85$\%$ and the pion misidentification rate was about 5$\%$.

     As with the $\Lm$ sample, the $\Xim-\Xibar$ and $\Om-\Ombar$ invariant
mass plots were fit to a Gaussian signal plus linear background for each
interval of $\xf$ and $\pt$. The total reconstructed mass distributions
are shown in Figs.~\ref{fig-masses}(c) through (f). With the final selection
criteria and the full E791 data set, we found 996\,180 $\pm$ 1\,200 $~\Xim$,
706\,620 $\pm$ 1\,020 $~\Xibar$, 8\,750 $\pm$ 110 $~\Om$, and 7\,460 $\pm$
100 $~\Ombar$ after background subtraction. Again, these numbers and their
errors come from the sum of signals in all bins. We checked that the $\Xim$
contamination for the $\Om$'s, after all cuts, was negligible.\\


      For each $\xf$ and $\pt$ bin and for each hyperon $Y$, we defined an
asymmetry parameter $A$ as
\begin{equation}
A\ysubb \equiv \frac{N\ysub - N\ybarsub ~r\ysub}
{N\ysub + N\ybarsub ~r\ysub} \; ~;~~ r\ysub = {\epsilon\ysub
\over \epsilon\ybarsub},
\label{eq1}
\end{equation}
where $N\ysub$ ($N\ybarsub$) is the number of hyperons (antihyperons)
produced in the bin, and $\epsilon\ysub$ ($\epsilon\ybarsub$), the
product of the geometrical acceptance and reconstruction efficiency for each
hyperon (antihyperon). Values for the $N$'s were obtained
from the individual fits to the mass plots for events selected to lie
within each $\xf$ and $\pt$ range.

\begin{figure}
\centerline{\epsfxsize=7.0in \epsffile{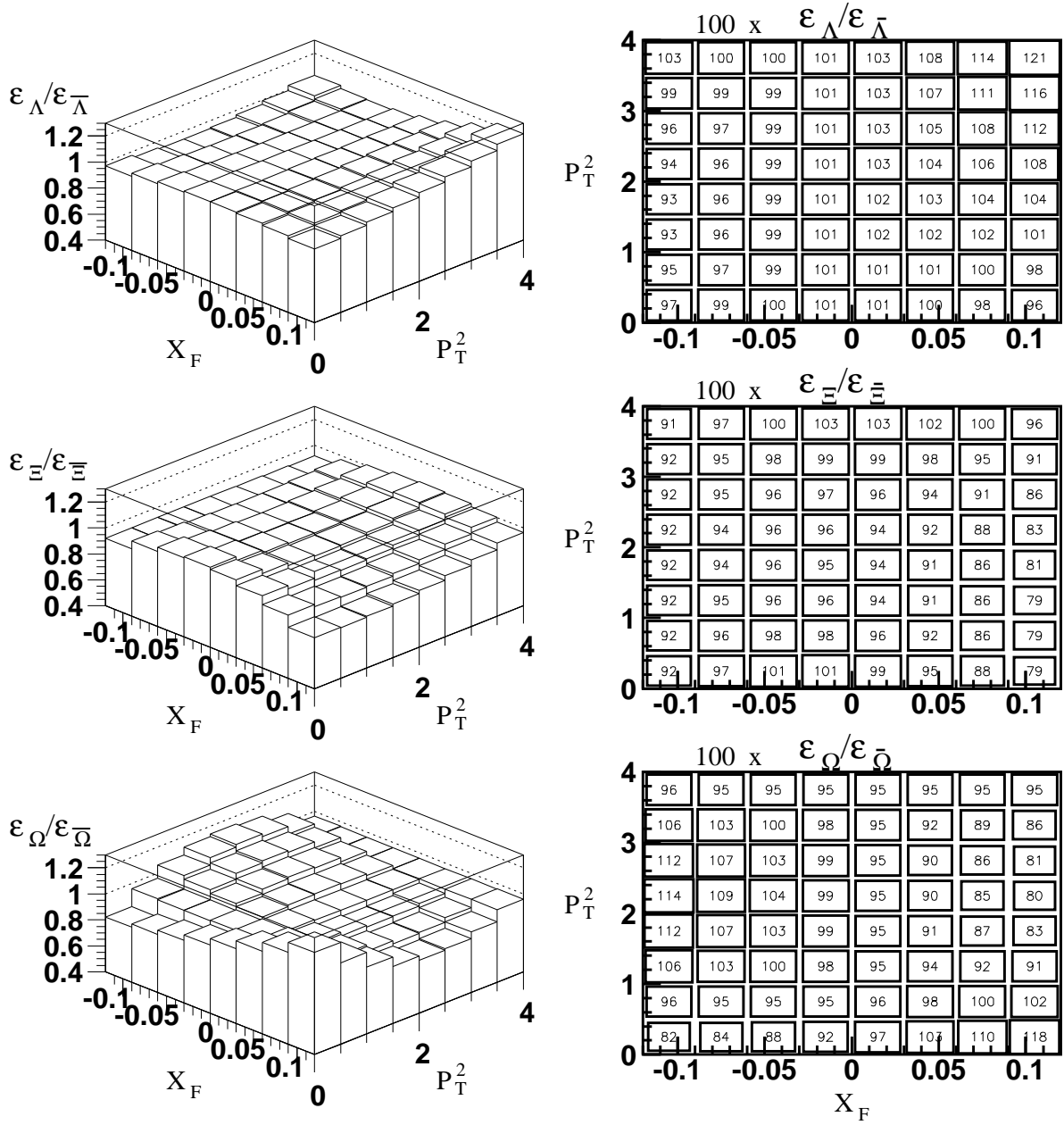}}
\caption{Ratio of efficiencies as functions of $\xf$ and $\pt$ for the three
hyperons. The transverse momentum axes are in units of (GeV/$c$)$^2$.}
\label{effxfpt_l0xiom}
\end{figure}

 Selection criteria for the particle and antiparticle samples were identical.
However, geometrical acceptances and reconstruction efficiencies were not
necessarily the same, mostly due to an inefficient region in the drift chambers
produced by the negative pion beam. To evaluate this effect, a large
sample of Monte Carlo (MC) events was created using the \PYTHIA/\JETSET~event
generators \cite{pyth}. These were projected through a detailed simulation of
the E791 spectrometer to simulate ``data'' in digitized format which was then
processed through the same computer reconstruction code as that used for data
from the detector. Candidate events were then subjected to the same selection
criteria as that used for data. To account for correlations between $\xf$ and
$\pt$, efficiencies were determined in bins of the two parameters.
The ratios of efficiencies $\epsilon\ysub/\epsilon\ybarsub$ $(\xf,\pt)$
evaluated from the MC samples are shown for each hyperon in
Figs.~\ref{effxfpt_l0xiom}(a)--(c). For the $\xf$ range of our data, there are
only minor differences between particle and antiparticle efficiencies. This
result was expected, as hyperon decays with tracks aimed at
the inefficient part of the drift chambers were rare. As a further check, we
 verified that our results were not significantly affected by symmetrically
eliminating these events.

\begin{figure}
\centerline{\epsfxsize=7.0in \epsffile{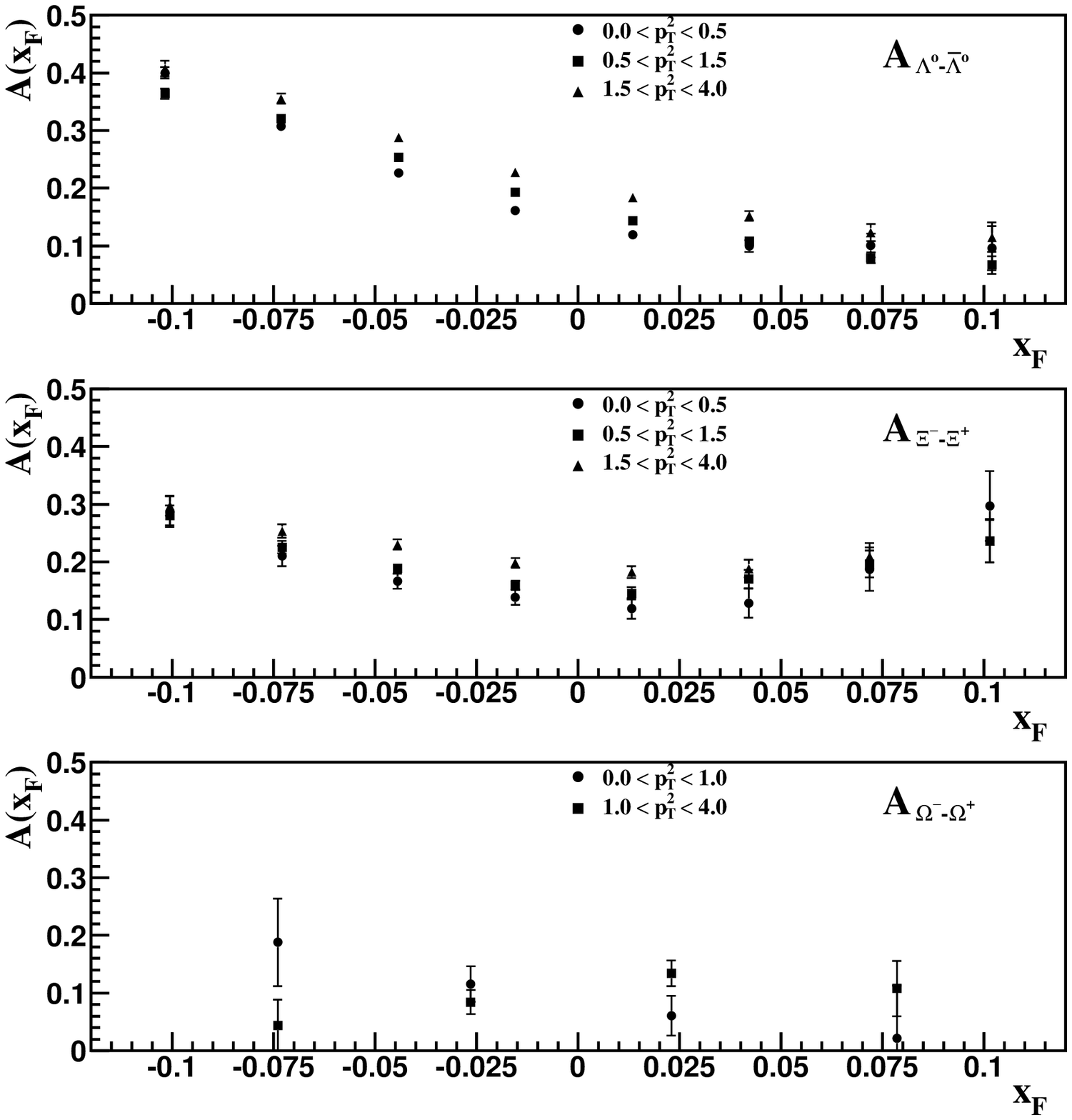}}
\caption{$\Lm-\Lbar$ (upper), $\Xim-\Xibar$ (middle), and
$\Om-\Ombar$ (lower) production asymmetries as a function of $\xf$
in different $\pt$ regions.}
\label{asy_xf_l0xiom}
\end{figure}

\begin{figure}
\centerline{\epsfxsize=7.0in \epsffile{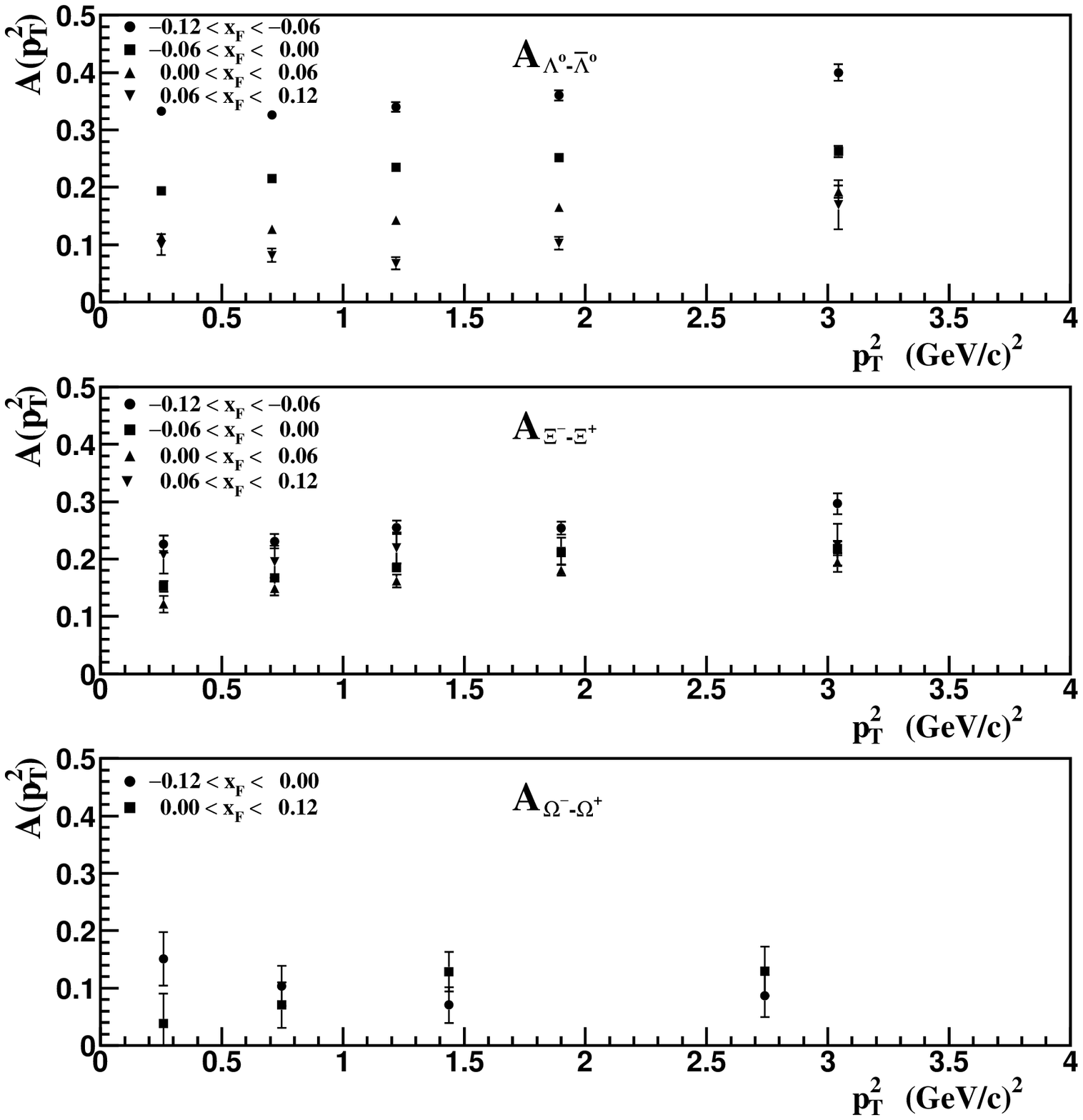}}
\caption{$\Lm-\Lbar$ (upper), $\Xim-\Xibar$ (middle), and
$\Om-\Ombar$ (lower) production asymmetries as a function of
$\pt$ in different $\xf$ regions.}
\label{asy_pt_l0xiom}
\end{figure}

  The asymmetries obtained using Eq.~\ref{eq1}
are shown in Figs.~\ref{asy_xf_l0xiom} and \ref{asy_pt_l0xiom} for
the hyperons $\Lm$, $\Xim$, and $\Om$, as a function of $\xf$ and
$\pt$, respectively. For each hyperon the asymmetry is presented as functions
of $\xf$ for different intervals of $\pt$, and also as functions of $\pt$ for
different intervals of $\xf$. The asymmetries show a substantial dependence
on $\xf$ and some dependence on $\pt$. There is evidence for a correlation
between $\xf$ and $\pt$.

\begin{figure}
\centerline{\epsfxsize=7.0in \epsffile{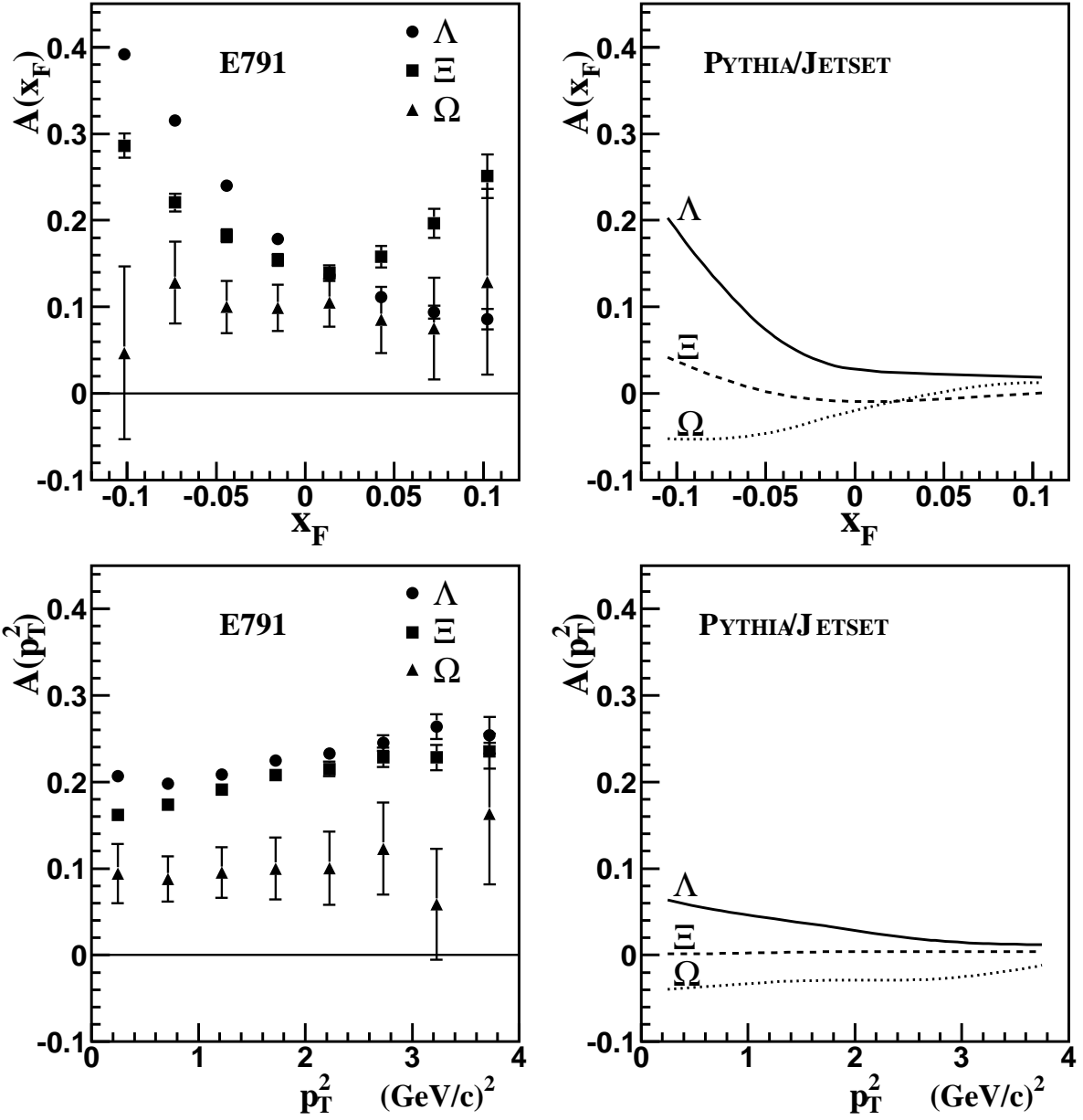}}
\caption{Comparison among the asymmetries in the production of the three
hyperons for the E791 data and \PYTHIA/\JETSET ~predictions. The error bars
of the E791 data include the statistical and systematic uncertainties}
\label{testnxfpt}
\end{figure}

   The asymmetry $A(\xf)$ integrated over $\pt$, and the asymmetry $A(\pt)$
integrated over $\xf$, are presented in Fig.~\ref{testnxfpt}. This figure
shows asymmetries for all three types of hyperons.
The results are also listed in Table~\ref{tb_asyc} along with statistical and
systematic errors.


We looked for systematic effects from the following sources:
\begin{itemize}
\item {Event selection criteria;}
\item {
The minimum transverse energy in the calorimeters required
in the event trigger;}
\item {
Uncertainties in calculating relative efficiencies for
particle and antiparticle;}
\item {
Effect of the 2.5$\%$ $K^-$ contamination in the beam.}
\end{itemize}

The first two effects were found to be negligible. A significant error came
from the spectrometer efficiencies, and these uncertainties are included as
systematic errors in Table~\ref{tb_asyc}.

The effect of the $K^-$ contamination in the beam was difficult to estimate as no
data on hyperon asymmetries in $K^-$ production in this $\xf$ range existed.
However, even for equal $\pi^-$ and $K^-$ production of $\Lm$'s (as opposed to
$\Lbar$'s) and 100$\%$ asymmetry for kaons (i.e., no $\Lbar$ production),
only a 1.5$\%$ change would occur in the final result.
As for the $\pi^-$ ($\bar u d$) beam,
no rise in the asymmetry in the positive $\xf$ direction was expected since
both $\Lm (uds)$ and $\Lbar (\bar u \bar d \bar s)$ shared one valence quark
with the $K^-(\bar u s)$ beam. Similarly, any rise in the
$\Xim (dss)-\Xibar (\bar{d}\bar{s}\bar{s})$ asymmetry should have been approximately
the same as for a $\pi^-$ beam  since, in both cases, the $\Xim$ hyperon
shared one valence quark with the beam particle. Since the $\Om (sss)$
was a leading particle for a $K^-$ beam, in this case some asymmetry was
expected in the forward $\xf$ region. We noted the low $K^-$ content of
the beam and the low $\xf$ of our data. In any event, no effect was evident.
Thus we did not correct our measured values of asymmetry for possible effects
of kaon contamination in the beam.

The asymmetry curves for $\Lm$ and for $\Xim$ cross over at $\xf\sim 0.02$.
At larger positive $\xf$ the $\Xim$ asymmetry becomes considerably
larger than that for $\Lm$. The fact that both beam and target
particles share one valence quark with the $\Xim$ and none with the
$\Xibar$, would predict that the $\Xim$ asymmetries rise with increasing
$\left|\xf\right|$ in both positive and negative $\xf$
regions, as observed in our data. In contrast, the $\Lm$ shares two
valence quarks with the target and only one with the beam particles,
while the $\Lbar$ only shares one with the beam. Consequently the
$\Lm-\Lbar$ asymmetry rises sharply
for $\xf<0$ and not for $\xf>0$. The $\Om-\Ombar$ asymmetry appears to
be constant. Notice that neither the $\Om$ nor the $\Ombar$
shares valence quarks with either the beam or target particles.

The observed positive asymmetry for $\Omega$ could arise, in part, from the
associated production of kaons. This could also be the origin of the
observed positive asymmetry near $\xf=0.0$ for $\Xi$'s and $\Lambda$'s.
Part of the associated production enhancement of baryons, as opposed to
antibaryons, may come from the higher energy thresholds for the production of
antibaryons. For production of hyperons, the conservation of strangeness
requires only the associated production of one or more kaons. For production of
an antihyperon, baryon number as well as strangeness must be conserved,
requiring the associated production of at least two additional baryons, thus
raising the energy thresholds and favoring particle over antiparticle
production \cite{capella1}.

The behavior of the three asymmetries shown in Fig.~\ref{testnxfpt} gives
evidence for the leading particle effect. In the backward region ($\xf<0$),
a larger asymmetry is observed when there is a larger difference in the
number of valence quarks in common with the target, for hyperon and
antihyperon. Evidence for a similar effect in the production of $D^{\pm}$
mesons in the forward region ($\xf>0$) was presented by the E791
collaboration~\cite{charm-exp} and others~\cite{e769-wa82}. The E791
collaboration has also studied asymmetries in the production of
$D^{\pm}_{s}$
~\cite{sudeshna} and $\Lambda^{\pm}_{c}$~\cite{hyp99}. However, leading
particle effects for charmed hadrons typically occur at larger values
of $\xf$ than they do for strange hadrons.

   The \PYTHIA/\JETSET ~\cite{pyth} model describes only some features of
our results, and those only qualitatively, as can be seen in
Fig.~\ref{testnxfpt}. This model predicts small  values of asymmetry for
$\xf \sim 0$; this is in contrast with our results which range from 0.08
to 0.18 in this region. Our data show that, even in the central  region, the
asymmetries are not zero, and suggest that leading particle effects play an
even larger role than expected as $\left|\xf\right|$ increases.

 In summary we report the most precise, systematic study to date of the
production asymmetry for $\Lambda$, $\Xi$, and $\Omega$ hyperons in a single
experiment. The range of $\xf$ covered, $-0.12 \leq \xf \leq 0.12$, allows
the study of asymmetries in regions close to $\xf=0$ for the first time in a
fixed target experiment. Some evidence for possible correlations between $\xf$
and $\pt$ is observed (see Figs.~\ref{asy_xf_l0xiom} and
\ref{asy_pt_l0xiom}). Our  results for particle-antiparticle asymmetries are
consistent with the  experimental results obtained by previous
experiments~\cite{accmor1,accmor2} (see Table~\ref{tb_tot}) but with smaller
uncertainties.   Our results can be described qualitatively in terms of the
energy thresholds for the production of hyperons and antihyperons
together with their associated particles and a model in which
the recombination of valence and sea quarks in the beam and target
particles contributes to the hyperon and antihyperon production in
an asymmetrical manner~\cite{strange-mod}.

        We gratefully acknowledge the assistance of the staffs of Fermilab and
of all the participating institutions. This research was supported by the
Brazilian Conselho Nacional de Desenvolvimento Cient\'{\i}fico e Tecnol\'ogico,
CONACyT (Mexico), the U.S.-Israel Binational Science Foundation, the U.S.
National Science Foundation and the U.S. Department of Energy. Fermilab is
operated by the Universities Research Associates, Inc., under  contract with
the United States Department of Energy.

\newpage
\newpage
\newpage
\newpage
\newpage
\newpage
\begin{table}[v]
\tabcolsep=2mm
\centering
\caption{Final asymmetries of $\Lambda$, $\Xi$, and $\Omega$ showing the
statistical plus systematical errors. The statistical errors given include
those due to the number of observed events and the number of MC events.
Typically, the error is dominated by the MC uncertainty.}
\vskip 4pt
\begin{tabular}{r@{  }r cccc} \hline

\multicolumn{1}{c}{$\xf$ Region} & $A_{\Lm-\Lbar}=\frac{\Lm-\Lbar}{\Lm+\Lbar}$ &
$A_{\Xim-\Xibar}=\frac{\Xim-\Xibar}{\Xim+\Xibar}$ & $A_{\Om-\Ombar}=\frac{\Om-\Ombar}{\Om+\Ombar}$ \\ \hline
-0.12 -- -0.09~   & 0.392  $\pm$  0.007  $\pm$  0.001 & 0.286 $\pm$ 0.014 $\pm$ 0.001 & 0.047 $\pm$ 0.096 $\pm$ 0.028  \\

-0.09 -- -0.06~   & 0.315  $\pm$ 0.005  $\pm$  0.001 & 0.220 $\pm$ 0.010 $\pm$ 0.001 & 0.128 $\pm$ 0.046 $\pm$ 0.005  \\

-0.06 -- -0.03~   & 0.240  $\pm$  0.004  $\pm$  0.001 & 0.182 $\pm$ 0.007 $\pm$ 0.001 & 0.100 $\pm$ 0.030 $\pm$ 0.001  \\

-0.03 --  0.00~   & 0.179  $\pm$  0.002  $\pm$  0.001 & 0.154 $\pm$ 0.007 $\pm$ 0.001 & 0.099 $\pm$ 0.027 $\pm$ 0.002  \\

0.00 --  0.03~   & 0.136  $\pm$  0.002  $\pm$ 0.001 & 0.139 $\pm$ 0.009 $\pm$ 0.001 & 0.106 $\pm$ 0.028 $\pm$ 0.001  \\

0.03 --  0.06~   & 0.111  $\pm$  0.005  $\pm$  0.001 & 0.158 $\pm$ 0.012 $\pm$ 0.001 & 0.085 $\pm$ 0.038 $\pm$ 0.005  \\

0.06 --  0.09~   & 0.094  $\pm$ 0.008  $\pm$  0.001 & 0.196 $\pm$ 0.017 $\pm$ 0.001 & 0.075 $\pm$ 0.058 $\pm$  0.011  \\

0.09 --  0.12~   & 0.086  $\pm$  0.012  $\pm$  0.003 & 0.250 $\pm$ 0.025 $\pm$ 0.001 & 0.129 $\pm$ 0.106 $\pm$ 0.018  \\ \hline

\multicolumn{1}{c}{$\pt$ Region} & $A_{\Lm-\Lbar}=\frac{\Lm-\Lbar}{\Lm+\Lbar}$ &
$A_{\Xim-\Xibar}=\frac{\Xim-\Xibar}{\Xim+\Xibar}$ & $A_{\Om-\Ombar}=\frac{\Om-\Ombar}{\Om+\Ombar}$ \\ \hline

 0.0 -- 0.5~   & 0.206  $\pm$  0.002  $\pm$  0.001 & 0.162 $\pm$ 0.007 $\pm$ 0.001 & 0.094 $\pm$ 0.034 $\pm$ 0.001  \\

0.5 -- 1.0~   & 0.198 $\pm$  0.002  $\pm$  0.001 & 0.173 $\pm$ 0.006 $\pm$ 0.001 & 0.088 $\pm$ 0.026 $\pm$ 0.001  \\

1.0 -- 1.5~   & 0.209  $\pm$  0.003  $\pm$  0.001 & 0.191 $\pm$ 0.005 $\pm$ 0.001 & 0.095 $\pm$ 0.029 $\pm$ 0.001  \\

1.5 -- 2.0~   & 0.224  $\pm$  0.003  $\pm$  0.001 & 0.207 $\pm$ 0.006 $\pm$ 0.001 & 0.099 $\pm$ 0.036 $\pm$ 0.001  \\

2.0 -- 2.5~   & 0.233  $\pm$  0.006 $\pm$  0.002 & 0.215 $\pm$ 0.008 $\pm$ 0.001 & 0.100 $\pm$ 0.042 $\pm$ 0.003 \\

2.5 -- 3.0~   & 0.245  $\pm$  0.008  $\pm$  0.005 & 0.229 $\pm$ 0.011 $\pm$ 0.001 & 0.123 $\pm$ 0.052 $\pm$ 0.011  \\

3.0 -- 3.5~   & 0.264 $\pm$  0.011  $\pm$  0.009 & 0.228 $\pm$ 0.015 $\pm$ 0.001 & 0.080 $\pm$ 0.062 $\pm$ 0.017  \\

3.5 -- 4.0~   & 0.254  $\pm$  0.015  $\pm$  0.015 & 0.235 $\pm$ 0.020 $\pm$ 0.001 & 0.163 $\pm$ 0.080 $\pm$ 0.019  \\ \hline

\end{tabular}
\label{tb_asyc}

\end{table}

\begin{table}[v]

\tabcolsep=2mm
\centering

\caption{Fully corrected production asymmetries of $\Lambda$,
$\Xi$, and  $\Omega$ integrated over three $\xf$ regions. Errors are obtained
by adding the  statistical and systematic errors in quadrature. Results
of ACCMOR~\cite{accmor1,accmor2} integrated over the range $0 \leq \xf \leq
0.35$ are shown for comparison.}

\begin{tabular}{lllll} \hline

& E791 & E791 & E791 & ACCMOR \\

& $-0.12\leq \xf \leq 0.12$ & $-0.12 \leq \xf
\leq 0$ &
 $0 \leq \xf \leq 0.12$ & $0 \leq \xf \leq 0.35$ \\ \hline

$A_{\Lm-\Lbar}$ & $0.207 \pm 0.001$ &
$0.242 \pm 0.002$ & $0.124 \pm 0.002$ & $0.119 \pm 0.009$~\cite{accmor1}\\

$A_{\Xim-\Xibar}$ & $0.176 \pm 0.004$ &
$0.186 \pm 0.004$ & $0.150 \pm 0.007$ & $0.130 \pm 0.050$~\cite{accmor2}\\

$A_{\Om-\Ombar}$ & $0.099 \pm 0.013$ &
$0.098 \pm 0.018$ & $0.100 \pm 0.021$ & $0.107 \pm 0.070$~\cite{accmor2}\\
\hline
\end{tabular}
\label{tb_tot}
\end{table}

\end{document}

%% file: author.tex
\author[inst_9]{E.~M.~Aitala,}
\author[inst_1]{S.~Amato,}
\author[inst_1]{J.~C.~Anjos,}
\author[inst_5]{J.~A.~Appel,}
\author[inst_14]{D.~Ashery,}
\author[inst_5]{S.~Banerjee,}
\author[inst_1]{I.~Bediaga,}
\author[inst_8]{G.~Blaylock,}
\author[inst_15]{S.~B.~Bracker,}
\author[inst_13]{P.~R.~Burchat,}
\author[inst_6]{R.~A.~Burnstein,}
\author[inst_5]{T.~Carter,}
\author[inst_1]{H.~S.~Carvalho,}
\author[inst_12]{N.~K.~Copty,}
\author[inst_9]{L.~M.~Cremaldi,}
\author[inst_18]{C.~Darling,}
\author[inst_5]{K.~Denisenko,}
\author[inst_3]{S.~Devmal,}
\author[inst_11]{A.~Fernandez,}
\author[inst_12]{G.~F.~Fox,}
\author[inst_2]{P.~Gagnon,}
\author[inst_1]{C.~Gobel,}
\author[inst_9]{K.~Gounder,}
\author[inst_5]{A.~M.~Halling,}
\author[inst_4]{G.~Herrera,}
\author[inst_14]{G.~Hurvits,}
\author[inst_5]{C.~James,}
\author[inst_6]{P.~A.~Kasper,}
\author[inst_5]{S.~Kwan,}
\author[inst_12]{D.~C.~Langs,}
\author[inst_2]{J.~Leslie,}
\author[inst_5]{B.~Lundberg,}
\author[inst_1]{J.~Magnin,}
\author[inst_14]{S.~MayTal-Beck,}
\author[inst_3]{B.~Meadows,}
\author[inst_1]{J.~R.~T.~de~Mello~Neto,}
\author[inst_16]{R.~H.~Milburn,}
\author[inst_1]{J.~M.~de~Miranda,}
\author[inst_16]{A.~Napier,}
\author[inst_7]{A.~Nguyen,}
\author[inst_3,inst_11]{A.~B.~d'Oliveira,}
\author[inst_2]{K.~O'Shaughnessy,}
\author[inst_6]{K.~C.~Peng,}
\author[inst_3]{L.~P.~Perera,}
\author[inst_12]{M.~V.~Purohit,}
\author[inst_9]{B.~Quinn,}
\author[inst_17]{S.~Radeztsky,}
\author[inst_9]{A.~Rafatian,}
\author[inst_7]{N.~W.~Reay,}
\author[inst_9]{J.~J.~Reidy,}
\author[inst_1]{A.~C.~dos Reis,}
\author[inst_6]{H.~A.~Rubin,}
\author[inst_9]{D.~A.~Sanders,}
\author[inst_3]{A.~K.~S.~Santha,}
\author[inst_1]{A.~F.~S.~Santoro,}
\author[inst_3]{A.~J.~Schwartz,}
\author[inst_4,inst_17]{M.~Sheaff,}
\author[inst_7]{R.~A.~Sidwell,}
\author[inst_1]{F.~R.~A.~Sim\~ao,}
\author[inst_18]{A.~J.~Slaughter,}
\author[inst_3]{M.~D.~Sokoloff,}
\author[inst_1]{J.~Solano,}
\author[inst_7]{N.~R.~Stanton,}
\author[inst_17]{K.~Stenson,}
\author[inst_9]{D.~J.~Summers,}
\author[inst_18]{S.~Takach,}
\author[inst_5]{K.~Thorne,}
\author[inst_7]{A.~K.~Tripathi,}
\author[inst_17]{S.~Watanabe,}
\author[inst_14]{R.~Weiss-Babai,}
\author[inst_10]{J.~Wiener,}
\author[inst_7]{N.~Witchey,}
\author[inst_18]{E.~Wolin,}
\author[inst_9]{D.~Yi,}
\author[inst_7]{S. Yoshida,}
\author[inst_13]{R.~Zaliznyak,}
\author[inst_7]{C.~Zhang}

\address[inst_1]{Centro Brasileiro de Pesquisas F{\'\i}sicas, Rio de Janeiro,
                 Brazil}
\address[inst_2]{University of California, Santa Cruz, California 95064, USA}
\address[inst_3]{University of Cincinnati, Cincinnati, Ohio 45221, USA}
\address[inst_4]{CINVESTAV, Mexico}
\address[inst_5]{Fermilab, Batavia, Illinois 60510, USA}
\address[inst_6]{Illinois Institute of Technology, Chicago, Illinois 60616, USA}
\address[inst_7]{Kansas State University, Manhattan, Kansas 66506, USA}
\address[inst_8]{University of Massachusetts, Amherst, Massachusetts 01003, USA}
\address[inst_9]{University of Mississippi--Oxford, University, MS 38677, USA}
\address[inst_10]{Princeton University, Princeton, New Jersey 08544, USA}
\address[inst_11]{Universidad Autonoma de Puebla, Mexico}
\address[inst_12]{University of South Carolina, Columbia, South Carolina 29208, USA}
\address[inst_13]{Stanford University, Stanford, California 94305, USA}
\address[inst_14]{Tel Aviv University, Tel Aviv 69978, Israel}
\address[inst_15]{Box 1290, Enderby, British Columbia V0E 1VO, Canada}
\address[inst_16]{Tufts University, Medford, Massachusetts 02155, USA}
\address[inst_17]{University of Wisconsin, Madison, Wisconsin 53706, USA}
\address[inst_18]{Yale University, New Haven, Connecticut 06511, USA}